\documentclass[conference]{IEEEtran}
\IEEEoverridecommandlockouts
\usepackage{cite}
\usepackage{amsmath,amssymb,amsfonts}
\usepackage{algorithmic}
\usepackage{graphicx}
\usepackage{textcomp}
\usepackage{xcolor}
\usepackage{caption}
\usepackage{subcaption}

\usepackage{epsfig,amsfonts,cases,graphics, latexsym, times, algorithmic, algorithm, bbm, color, soul,wrapfig,comment}
\def\BibTeX{{\rm B\kern-.05em{\sc i\kern-.025em b}\kern-.08em
    T\kern-.1667em\lower.7ex\hbox{E}\kern-.125emX}}
\begin{document}

\title{Model Predictive Control approach to adaptive messaging intervention for physical activity  \\
\thanks{This work has been partially supported by National Institutes of Health (NIH) Grant R01 HL142732 and National Science Foundation (NSF) Grant 1808266.}
}

\author{\IEEEauthorblockN{Ibrahim E. Bardakci}
\IEEEauthorblockA{\textit{School of Electrical Engineering \& Computer Science} \\
\textit{The Pennsylvania State University}\\
University Park, USA \\
ibrahimekrem@gmail.com}
\and
\IEEEauthorblockN{Sahar Hojjatinia}
\IEEEauthorblockA{\textit{School of Electrical Engineering \& Computer Science} \\
\textit{The Pennsylvania State University}\\
University Park, USA \\
suh434@psu.edu}
\and
\IEEEauthorblockN{Sarah Hojjatinia}
\IEEEauthorblockA{\textit{AS \& UX} \\
\textit{Aptiv}\\
Troy, USA \\
sarah.hojjatinia@gmail.com}
\and
\IEEEauthorblockN{Constantino M. Lagoa}
\IEEEauthorblockA{\textit{School of Electrical Engineering \& Computer Science} \\
\textit{The Pennsylvania State University}\\
University Park, USA \\
cml18@psu.edu}
\and
\IEEEauthorblockN{David E. Conroy}
\IEEEauthorblockA{\textit{Department of Kinesiology} \\
\textit{The Pennsylvania State University}\\
University Park, USA \\
conroy@psu.edu}
}

\maketitle

\begin{abstract}
In this work, we have developed a framework for synthesizing data driven controllers for a class of uncertain switched systems arising in an application to physical activity interventions. In particular, we present an application of probabilistic model predictive control to design an efficient, tractable, and adaptive intervention using behavioral data sets i.e. physical activity behavior. The models of physical activity for each individual are provided for the design of controllers that maximize the probability of achieving a desired physical activity goal subject to intervention specifications. We have tailored the mixed-integer programming-based approach for evaluating the Model Predictive Control decision at each time step. 
\end{abstract}


\section{Introduction}
Physical inactivity is recognized as one of the major risk factors for many chronic diseases, including diabetes, cardiovascular diseases, stroke, and different types of cancer. It has also been identified as the fourth leading risk factor for global mortality ($6\%$ of deaths globally)~\cite{world2010global}. While physical activity is widely recommended and is known to have multi-system benefits for health, most American adults fail to achieve recommended levels of physical activity~\cite{Bennie2019}. Given the economic burden of physical inactivity on one hand and health issues caused by it on the other hand, intervention programs to increase physical activity and reduce the sedentary behavior are needed. 

With the prevalence of smart phones and activity monitors, an effective method to increase physical activity has shown to be message interventions~\cite{smith2020text}. The fact that physical activity is a dynamic behavior and can be influenced by external factors such as weather conditions should be considered in interpreting the outcomes of the intervention~\cite{turrisi2021seasons}. Personalized dynamical models of physical activity based on message interventions developed by our group have shown a positive effect on increasing physical activity level~\cite{conroy2019personalized, dosefinding21}. However, the change in physical activity differs for each individual as a function of the type of intervention message, temperature, precipitation, and day of week.
These heterogeneous responses of people to different types of interventions a need for designing and implementing more efficient interventions, adaptive to each individual's behavior and current context that balance the potential for achieving behavioral goals against the threat of disengagement from excessive intervention burden, as well as reducing the burden treatment.

In this regard, the objective of this paper is to develop a computationally tractable framework for synthesizing data-driven controllers for a class of uncertain switched systems arising in an application to increase physical activity level in individuals. We present a design of personalized behavior monitoring systems using control systems tools, aimed at maintaining desired levels of physical activity.
In this regard, we use personalized models of physical activity described by a discrete-time dynamical system of the form
\begin{equation}\label{mod1}
	y_{k} \triangleq f(y_{k-1},y_{k-2},\hdots,y_{k-n},u_{k},\hdots,u_{k-n},w_k)
\end{equation}
for $ k=1,\hdots,T $ where $ y_k \in \mathcal{Y}$ is the system output, $ u_k \in \mathcal{U} $ is the control input and $ w_k \in \mathcal{W} $ is the process noise or disturbance, all at time step $ k $. The function $ f: \mathcal{Y}\times\mathcal{U}\times\mathcal{W} \rightarrow \mathcal{Y} $ are the input-output transition functions. The process noise trajectory $ (w_1, w_{2},\hdots,w_{T}) \in \mathcal{W}$ is random, with a known distribution. In equation~\eqref{mod1}, the state $ y_k $ at time $ k $ is a function of $  n $ previous values of the state (order of the model), inputs $ u^j_k \in \mathcal{U} $, and uncertainties $ w_k \in \mathcal{W} $ at time k. These dynamical systems that can model the physical activity behavior of an individual are provided as a basis for developing controllers to maintain optimal levels of physical activity volume. 

In this paper, we use the developed model representing the dynamical behavior of a subject's physical activity as a Picewise Affine (PWA) model. The dynamical models of physical activity used in this paper were developed based on the data recorded from the participants recruited for a 6-month messaging intervention study with the aim to increase their physical activity \cite{conroy2019personalized, dosefinding21}. A PWA model is obtained by partitioning the state space into a finite number of regions and associating each of these regions with an affine sub-model. These sub-models describe possible continuous dynamics, and are coupled via a discrete logic that determines which subsystem is active at each time instant. PWA models are tractable approximations of many non-linear phenomena~\cite{sontag1981nonlinear} and they can be used to approximate the behavior of any complex dynamical system~\cite{bemporad2005bounded}. The PWA model to be used for representing the physical activity of a subject has sub-models of the form
\begin{equation}\label{PWA}
	y_k = a^{\sigma(k)}_0+\sum_{i=1}^{n}a^{\sigma(k)}_{i}y_{k-i}+\sum_{j=1}^{m}\sum_{i=0}^{n}b^{\sigma(k)}_{ij}u^{j}_{k-i}+w_k,
\end{equation}
where the index $ \sigma(k) $ denotes the sub-model active at time $ k $, $ u^j_k $ indicates what type of message is sent at time k (if any), and $ w_k $ is noise. The active sub-model at each time $ k $ is determined based on the current and past states of the subject and external factors such as weather conditions, and day of week. The larger the number of sub-models and order of them, the more complex the PWA model becomes. The sub-models provide information on controller design to deal with non-linear moderators of intervention effects since they represent different intervention responses under different conditions.


Drawing on these personalized models of physical activity, the objective is to design controllers that maximize the probability of achieving a desired physical activity goal subject to intervention specifications. The effective operating regions for these controllers are specified by the intervention window. For the physical activity controller, the intervention window is defined as the times when participants are likely to respond to intervention, and it is obtained through the feedback in the modeling phase. The source of feedback typically comes from one of two sources: self-monitoring or sensors. Sensors embedded in wearable and portable devices have emerged as an attractive alternative to self-monitoring because they provide more intensive data, participant burden, and are not vulnerable to the same memory biases as self-monitoring~\cite{de2016activity}.

\subsection{Probabilistic model predictive control}
Once the identified model is given, a controller repeatedly selects when and what type of intervention to apply in a given window. Given the estimated distribution of the noise $w_k $ and information on the distributions of the external factors, at each time $ k $ the controller determines a sequence of interventions that maximizes the probability of achieving the goal while avoiding undesirable situations such as overburden from applying too many interventions. Hence, at every time instant, given a look ahead horizon $ T $, the controller solves an optimization problem of the form for $ k=1,2,\hdots,T $,
\begin{equation} \label{prob_max_mpc}
	\begin{aligned}
		\mbox{maximize} \qquad &\mbox{Prob} \left\{  \mbox{Desired Change in Physical Activity} \right\}\\
		\text{subject to}   \qquad   &  \mbox{Model }\&\mbox{ Intervention Constraints} 
	\end{aligned}
\end{equation}
However, this decision process of repeated interventions is highly complex. Therefore, in this study, we use a Model Predictive Control (MPC) approach as a way to design the intervention algorithm since model predictive control can efficiently handle  such complexities~\cite{morari1999model,mayne2014model, mesbah2016stochastic} and has successfully implemented in intervention designs~\cite{lagoa2014designing,zafra2010risk,nandola2010novel}. In the MPC problem, the first step is predicting future behavior of the system to be controlled using mathematical model of the system along with uncertainty descriptions. The second step consists of determining the sequence of control actions that is optimal in terms of specified performance measure (e.g., maximizing the desired objective or minimizing the cost). Finally, the first value of these control actions applies, then this whole process repeats with the new measurements. In the standard MPC, the horizon slides forward and for this reason MPC is also called receding horizon control. However, because of the nature of this intervention design, the intervention window is predefined. In other words, the horizon of MPC controller is reduced by one step at each time instant and is referred as a shrinking horizon MPC~\cite{skaf2010shrinking}.

The approach proposed here is aimed at providing intervention design procedures that can handle large amounts of stochastic uncertainty. Each step of MPC algorithm consists of probability maximization problem or chance-constrained optimization involving integer variables.

In our approach, we consider a sample-average approximation of the probability maximization problem~\eqref{prob_max_mpc} where the probability function in the objective approximated by a sample average estimates using Monte Carlo simulation. The resulting problem involving a discrete distribution can be difficult to solve~\cite{luedtke2010integer}. Moreover, the solutions obtained by these methods can be quite conservative. However, we consider a finite (or shrinking) horizon MPC, therefore we only have a finite number of randomly chosen instances of possible outcomes of parameter values. Hence, the integer-programming approach can be good candidate as a solution of the MPC problem. Following the procedure proposed in~\cite{luedtke2010integer,ahmed09sample} we consider integer-programming based approaches for evaluating such an MPC control decision at each time step.

\section{Objectives and Constraints of the Intervention Design}
In this problem, given the input-output description of the model, we aim at finding an input policy $ (u^j_1,u^j_2,\hdots,u^j_T) $, i.e. when and which message to send, that maximizes the probability of achieving the desired number of step count over $ T - n $ periods subject to model and input constraints. 
\begin{subequations} \label{main_problem2}
	\begin{alignat}{4} 
		&\!\max_{u_{k^*}^j,...,u_{T}^j}        &  & \mathbb{P} \left\{  \sum_{k=1}^{k^{*}-1} y_k +\sum_{k^{*}}^{T} y_k \geq Goal \right\} &\quad& \\
		&\text{subject to} &      &     &      &      \nonumber\\
		&    &        &  y_{k} = f(y_{k-1},y_{k-2},... ,y_{k-n},u^j_{k},w_k) &      &\\
		&    &         &   u^j_k \in \mathcal{U}    &
	\end{alignat}
\end{subequations}
where $\mathcal{U}$ is the input constraint and the desired step count $ Goal $ is determined by the historical data collected. 
At each time period $ k $, the algorithm determines the control action $ u^j_k $. When the control action is taken at time $ k^* $, the previous measurements $ (y_1,\hdots,y_{k^{*}-1}) $ are known; hence, the uncertainties until time $ k^* $, $ (w_1,\hdots,w_{k^{*}-1}) $, are  zero. However, the current period and future outputs $ (y_k^{*},\hdots,y_T) $ are still to be predicted and the uncertainties $ (w_k^{*},\hdots,w_{T})$ are present in the model. 

\subsection{Constraints on interventions}
In this section, we quantify the constraints on the text message interventions (or control inputs). These constraints are aimed at addressing the intervention or treatment burden~\cite{eton2013systematic, heckman2015treatment}.
In the intervention design phase, we elaborate algorithms that constrain the frequency and number of interventions. Using the feedback of participants in the study, these constraints on the text message interventions are specified mathematically as the input constraint set $ \mathcal{U} $
\begin{equation}\label{input_set}
	\begin{aligned}
		\mathcal{U} \triangleq \Bigl\{&u^j_k \in \{0,1\}: \sum_{k}^{k+1}\sum_{j=1}^{m}u^{j}_k \leq 1,\ \sum_{k=1}^{T}\sum_{j=1}^{3}u^{j}_k \leq \alpha,\\
		&\sum_{k=1}^{T} \sum_{j=1}^{3}c_k u^j_k \leq \beta, \ j=1,...,m,\ k=1,\hdots, T \Bigr\}
	\end{aligned}
\end{equation}
where input $u^j_k$ can take values $ 0 $ or $ 1 $ representing either applying or not applying intervention at each time. Sum of two consecutive inputs is considered to be less than 1 preventing more than one intervention (of any types) within 30 minutes (two sampling time of 15 minutes). The number of interventions applied in the given period $ T $ (intervention window or horizon) is constrained to be less than or equal to $ \alpha $, that should  be predefined. Total cost of interventions is less than $ \beta $. Here, $ c_k $ represents the cost of applying intervention $ u^j_k $ at time $ k $ for $ j=1,\hdots, 3 $. 


\subsection{Determining the costs on interventions}
In this model, we use a progressive penalty function on the text message interventions (inputs). It is a composite based on the time of day (uniform and progressively lower cost for messages sent later in the day) and person-specific temporal patterning of physical activity (greater cost for messages at times when a person is typically more active). The cost function associated with the time of day is introduced to prevent pile-up of messages early in the day (depleting the bank of messages that might be sent later in the day). The cost function associated with temporal patterning of physical activity introduced as a constraint based on national data showing hourly patterns in physical activity. 

\begin{figure}[H]
	\centering
	\begin{subfigure}{0.49\linewidth}
		\centering
		\includegraphics[width=\linewidth]{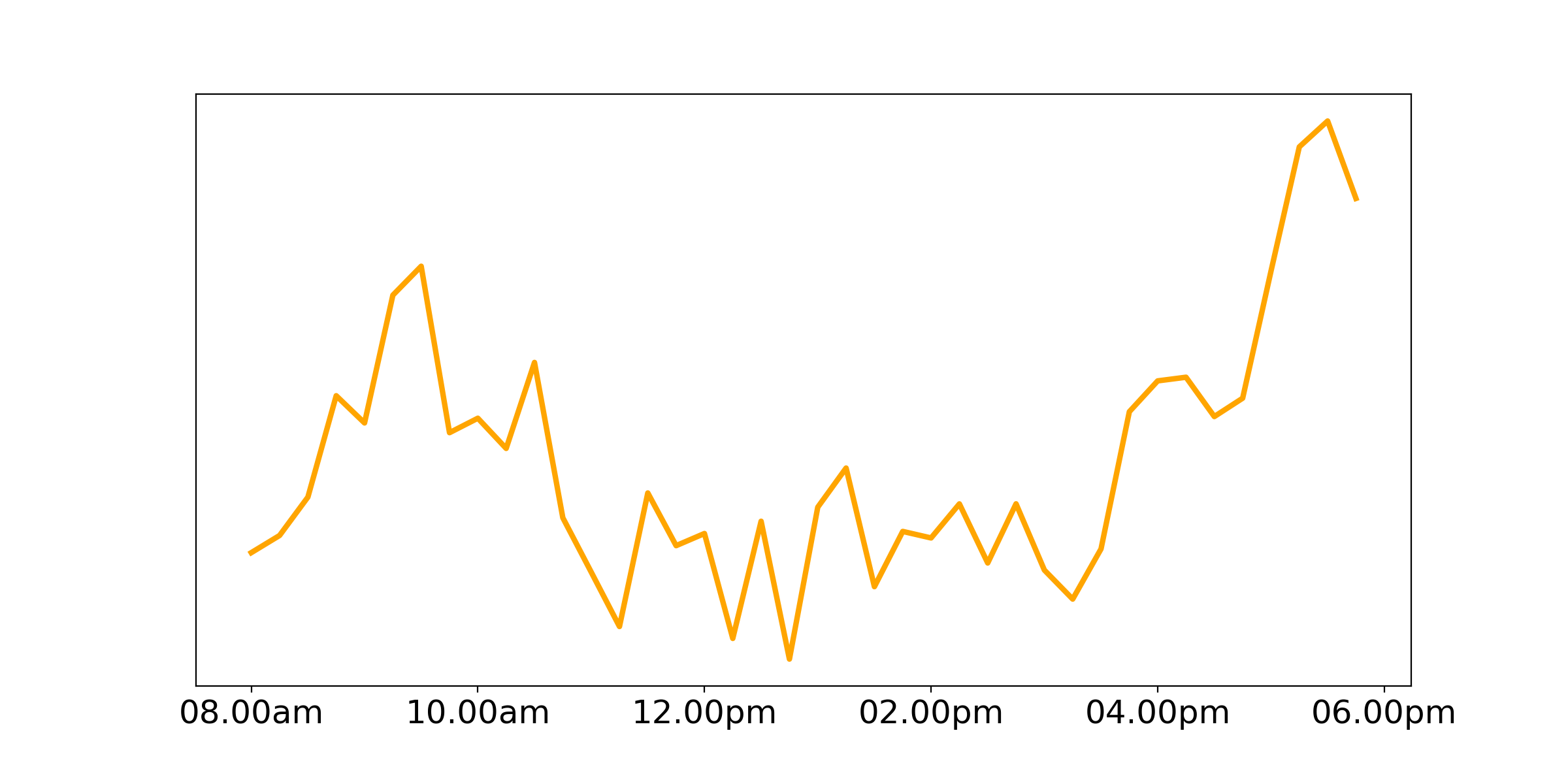}
		\caption{}
		\label{fig:fig:weight1} 
	\end{subfigure}
	\hfill
	\begin{subfigure}{0.49\linewidth}
		\centering
		\includegraphics[width=\linewidth]{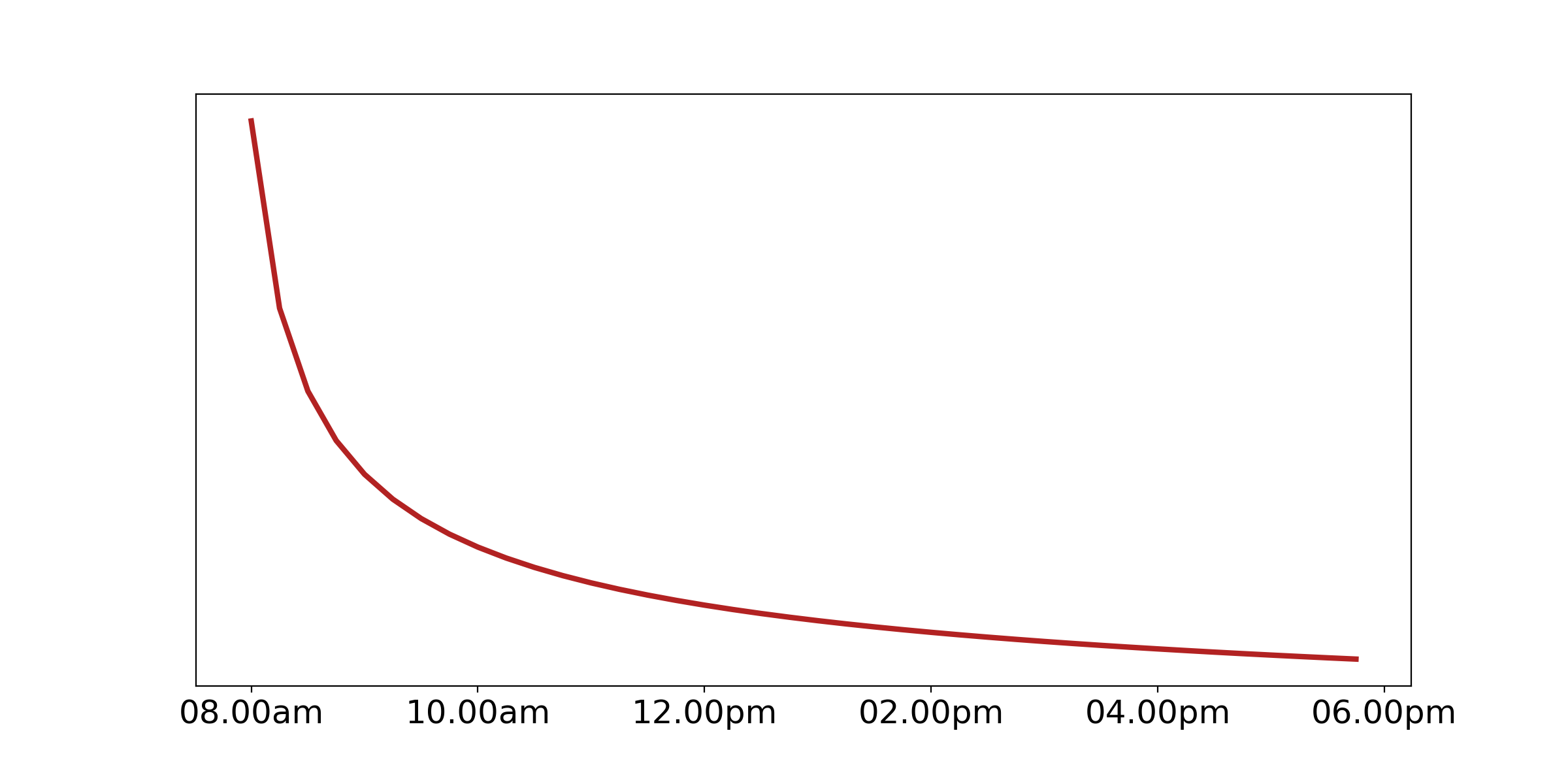}
		\caption{}
		\label{fig:weight2}
	\end{subfigure}
	\\[\baselineskip]
	\begin{subfigure}[H]{0.49\linewidth}
		\centering
		\includegraphics[width=\linewidth]{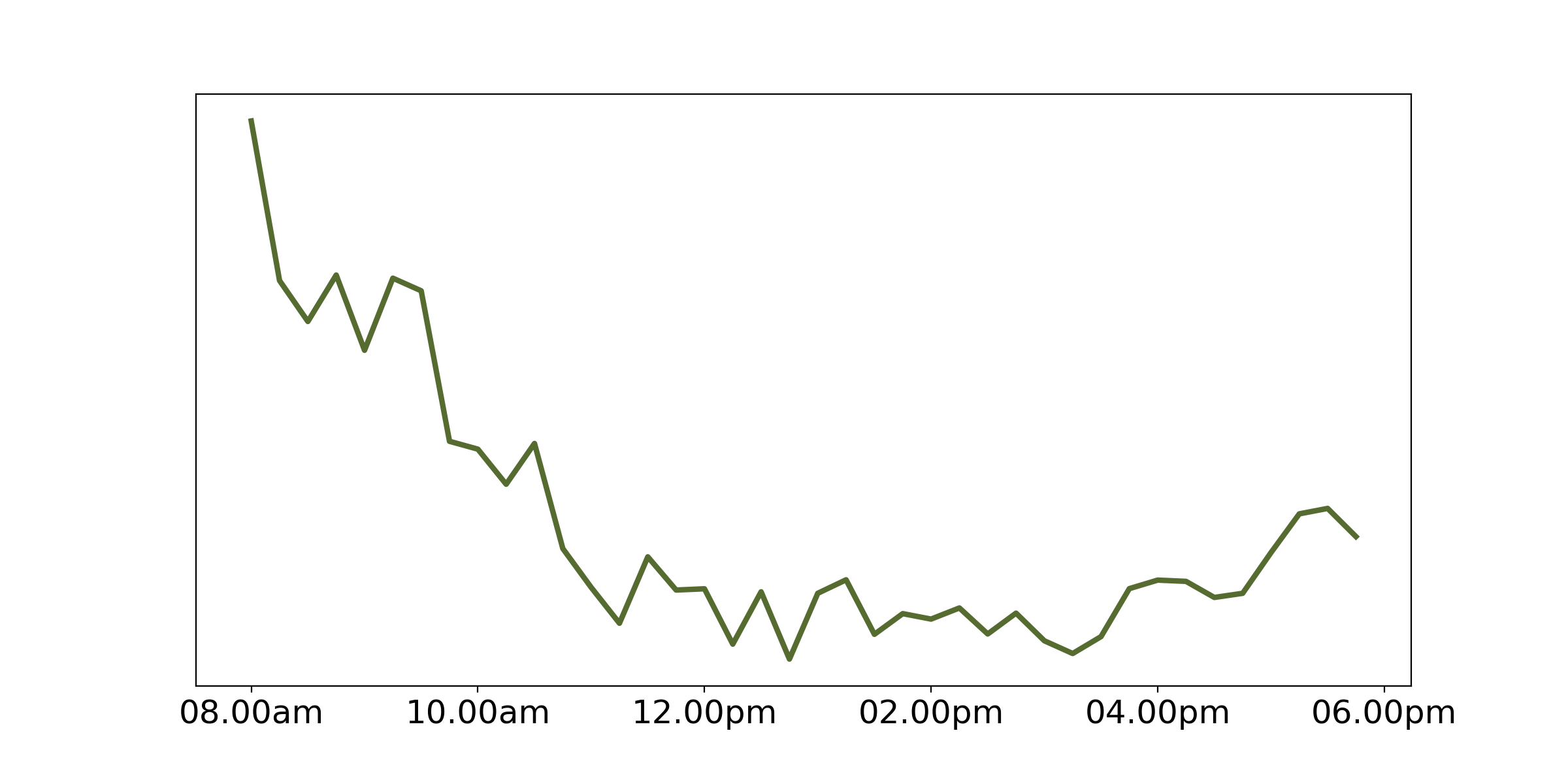}
		\caption{}
		\label{fig:comp_weight}
	\end{subfigure}
	\centering
	\caption{(a) Costs as a function of temporal patterning of physical activity ($ c^{step}_k $) (b) Costs as a function of time of the day ($ c^{time}_k $). (c) Composite costs  $ c_k $.}
	\label{fig:three-weights}
\end{figure}

More specifically we have sequence of cost functions $ c^{time}_k $ and $ c^{step}_k $ on input sequence $ u^t_k $ for $ k = n+1,\hdots,T  $. The cost function $ c^{step}_k $ is obtained from the historical data which is determined as the normalized hourly averages of step count of the participant for the specified intervention window (see Figure~\ref{fig:three-weights}-a). The function $ c^{time}_k $ is chosen as monotonically decreasing function of a time to reflect the cost of applying intervention later in the day is lower (see Figure~\ref{fig:three-weights}-b). Then, the composite cost function is obtained as $c_k = c^{time}_k \times c^{step}_k$ for $k=n+1,\hdots,T$.

\section{Stochastic Model Predictive Control for Physical Activity Intervention Design}
We now describe a general optimization problem and show how model predictive control can be implemented on the problem. At every time instant, given the last $ k $ measurements, the controller repeatedly solves the optimization problem \eqref{main_problem2}. Following the approach in~\cite{luedtke2010integer}, the problem~\eqref{main_problem2} can be formulated as the following stochastic mixed-integer program over the time interval $ k=n+1,n+2,\hdots,T $
\begin{subequations} 
	\begin{alignat*}{4}
		&\!\max       &\quad&\frac{1}{N}\sum_{s=1}^{N}p_s  & \tag{SMIP-MPC}& \\
		&\text{subject to} &      &     &      &      \\
		&    &      & \left( \sum_{k=1}^{k^{*}-1} y_k + \sum_{k=k^{*}}^{T} y^s_k - Goal\right) \geq M^s (p^s-1)  \qquad \qquad &    &\\
		&    &      &  y^s_k =a_0+\sum_{i=1}^{n}a_{i}y^s_{k-i}+\sum_{j=1}^{m}\sum_{i=0}^{n}b_{i j}u^{j}_{k-i}+w^s_k&  &\\
		&    &      &  p_s \in \{0,1\} &\\
		&    &      &  u^j_k \in \mathcal{U}  &
	\end{alignat*}
\end{subequations}
for $ s=1,\hdots,N  $ where $ M^s $ is a large positive number such that $M^s \geq \max_{u^j_k \in \mathcal{U}} \left( \sum_{k=1}^{k^{*}-1} y_k + \sum_{k=k^{*}}^{T} y^s_k - Goal\right)$ .

\begin{algorithm}[H] \caption{\bf Shrinking Horizon MPC algorithm}
	\em
	\label{algo:b}
	(0) Given the step count measurements $ y_1,\hdots,y_{k^{*}-1} $, past input sequence $ u^j_{1},u^j_{2},\hdots,u^j_{k^{*}-1} $, dynamical model $f$, desired step count Goal, the distribution of $ w_k $\\
	(1) Solve (SMIP-MPC) and obtain a future control sequence $ u_{k^{*}}^j,u_{k^{*}+1}^j,\hdots, u_T^j$ \\
	(2) Apply $ u_{k^{*}}^{j}$ as the current input.\\
	(3) Set $k^{*} \doteq k^{*}+1$, and return to (1), till $ k^{*}=T $
\end{algorithm}

More specifically, at time $k^{*} $, we first collect the measurements $ (y_{1},y_{2},\hdots,y_{k^{*}-1} )$, and past input sequence $( u^j_{1},u^j_{2},\hdots,u^j_{k^{*}-1} )$, and generate the current and future disturbance sequences $( w_k^{*}, w_{k^{*}+1},\hdots,w_{T} )$. Then, we solve the stochastic mixed-integer MPC problem over the remaining period, i.e., $k=k^{*},\hdots,T$, using the information given in the first step, and obtain the optimal current and future input sequence $ (u_{k^{*}}^j, u_{k^{*}+1}^j,\hdots,u^j_{T} )$. Finally, we use $ u_k^{j^*} $ as the current input. 

\section{Application to behavioral data sets}
In this section, we provide an implementation of the data driven approach to controller design for developing personalized treatments aimed at improving physical activity volume as measured by step counts.
Dynamic model of physical activity is provided as a 5-th order switched affine system with inputs being binary variables indicating if a message of any type was received or not, and output the current step counts. The model consists of two sub-systems based on weekdays and weekends. The process noise is normally distributed and $ \mu_w $ and $ \sigma_k $ are known for each time step $ k $. The PWA model of physical activity of an individual for weekdays is given as 
\begin{equation}\label{weekday_model}
	\begin{aligned}
		y_k = \ &80.51 - 0.0052y_{k-1} + 0.0043y_{k-2} + 0.0421y_{k-3}\\
		-&0.0674y_{k-4} +0.4607y_{k-5}\\
		-&20.418u^1_{k} + 33.621u^1_{k-1}  -9.370u^1_{k-2}  + 9.534u^1_{k-3}\\
		+&9.417u^1_{k-4} +  4.002u^1_{k-5}\\
		+&2.383u^2_{k} -4.976u^2_{k-1}  +5.695u^2_{k-2}+ 25.936u^2_{k-3}\\ -&1.737u^2_{k-4} +19.616u^2_{k-5}\\
		-&14.345u^3_{k}  +14.103u^3_{k-1} + 25.980u^3_{k-2}  +32.900u^3_{k-3}\\
		-&17.978u^3_{k-4} -61.739u^3_{k-5}
	\end{aligned}	
\end{equation}
where the disturbance $ w $ is assumed to be independent and normal with a mean $ \mu_w $ and standard deviation $ \sigma_w $. These parameters are also provided as a part of model identification phase given as $ \mu_w = -0.0155 $ and $ \sigma_w = 268.679 $. Since the model is a 5-th order affine system and sampling time is 15 minutes, inputs to the model are step count measurements and intervention history in the last 75 minutes and process noise at the current time.

The objective is to decide when and what type of messages to be sent (input) based on the identified switched model in order for the subject’s physical activity (state) to achieve the desired objectives. Then at each time point during the implementation of the intervention for a new individual, the developed algorithm uses the data collected on the individual to recommend an intervention for this individual at this time. Unlike the model given in~\eqref{weekday_model}, the controller requires step count measurements and intervention history from the beginning of intervention window and noise trajectory until the end of intervention window. The controller needs step count measurements as an input because the objective requires total number of step counts from $ k=1 $ to $ k=T $ where $ T $ is the final period of intervention window. Since the  interventions are subject to the constraints such as frequency and maximum number of interventions in the intervention window, the necessity for intervention history is clear. Given the model above, we use the Algorithm 1, to solve this optimization problem.

\begin{figure}
	\centering
	\begin{subfigure}{0.85\linewidth}
		\centering
		\includegraphics[width=\linewidth]{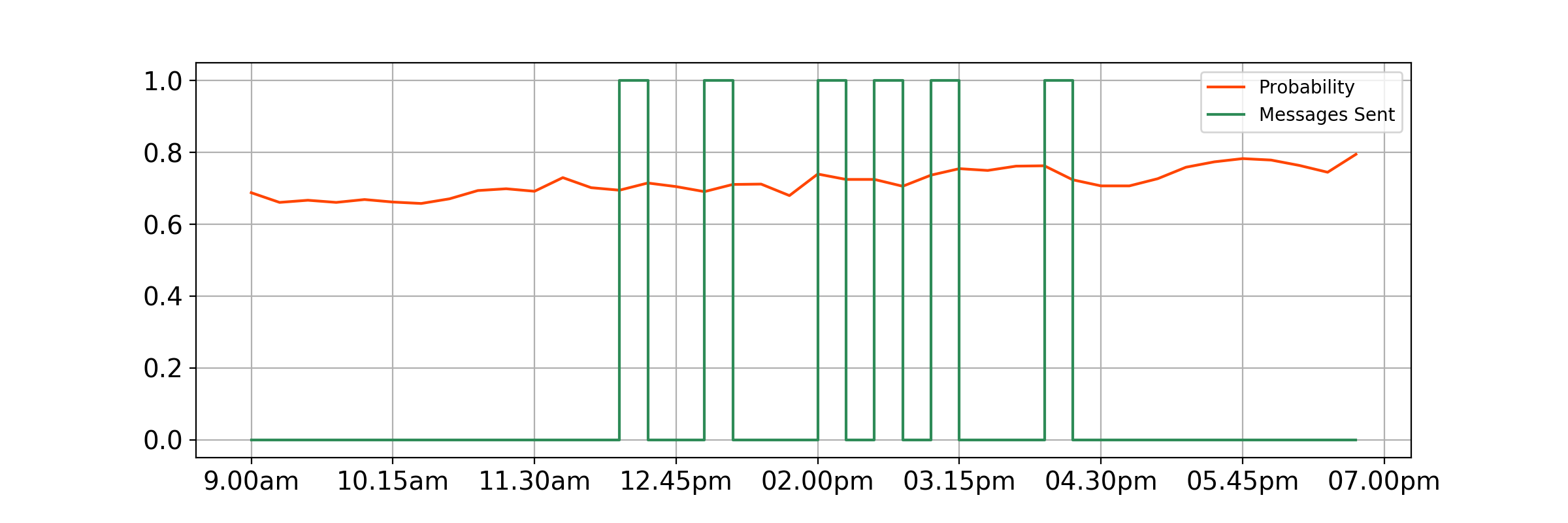}
		\caption{}
		\label{fig:input_weekday_normal} 
	\end{subfigure}
	\hfill
	\begin{subfigure}{0.85\linewidth}
		\centering
		\includegraphics[width=\linewidth]{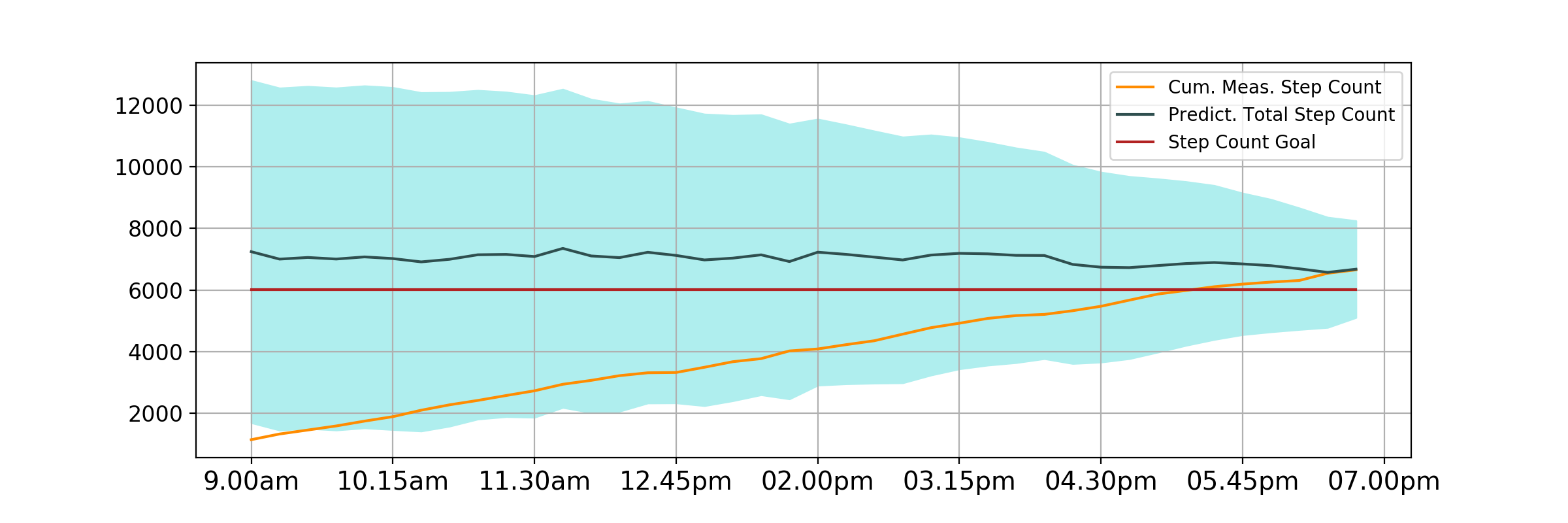}
		\caption{}
		\label{fig:detailed_weekday_normal}
	\end{subfigure}
	\centering
	\caption{Text message interventions and the probability of achieving the desired goal for regular physical activity }
	\label{fig:regular}
\end{figure}
\begin{figure}
	\centering
	\begin{subfigure}{0.9\linewidth}
		\centering
		\includegraphics[width=\linewidth]{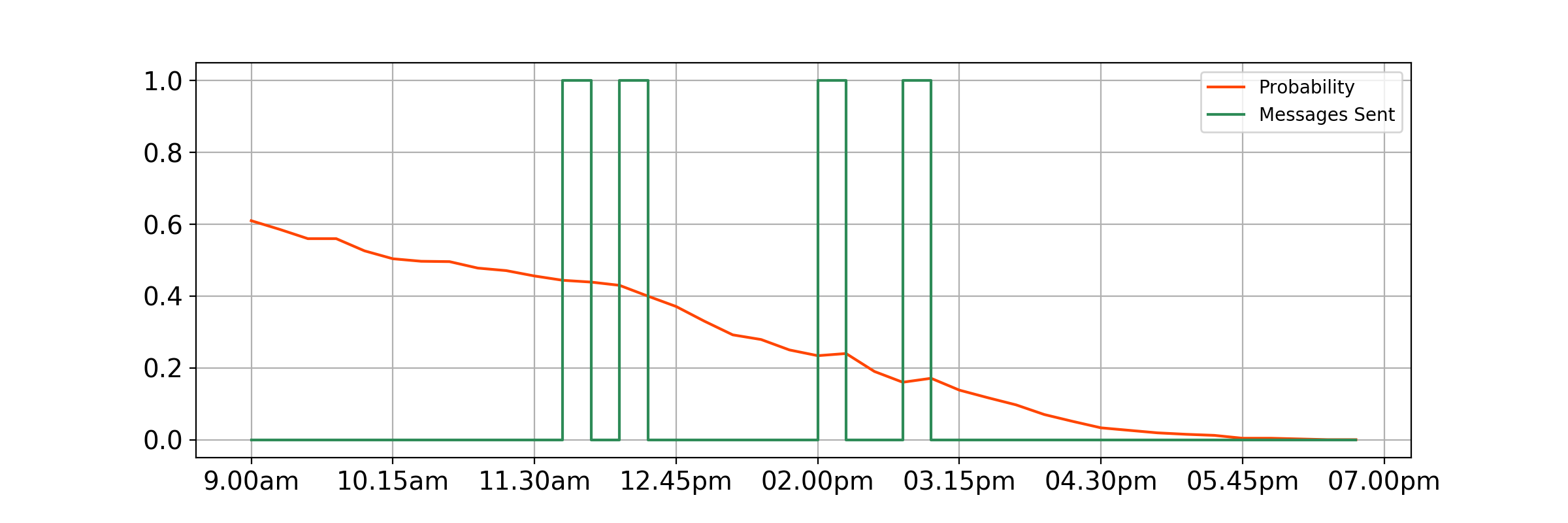}
		\caption{}
		\label{fig:input_weekday_low} 
	\end{subfigure}
	\hfill
	\begin{subfigure}{0.9\linewidth}
		\centering
		\includegraphics[width=\linewidth]{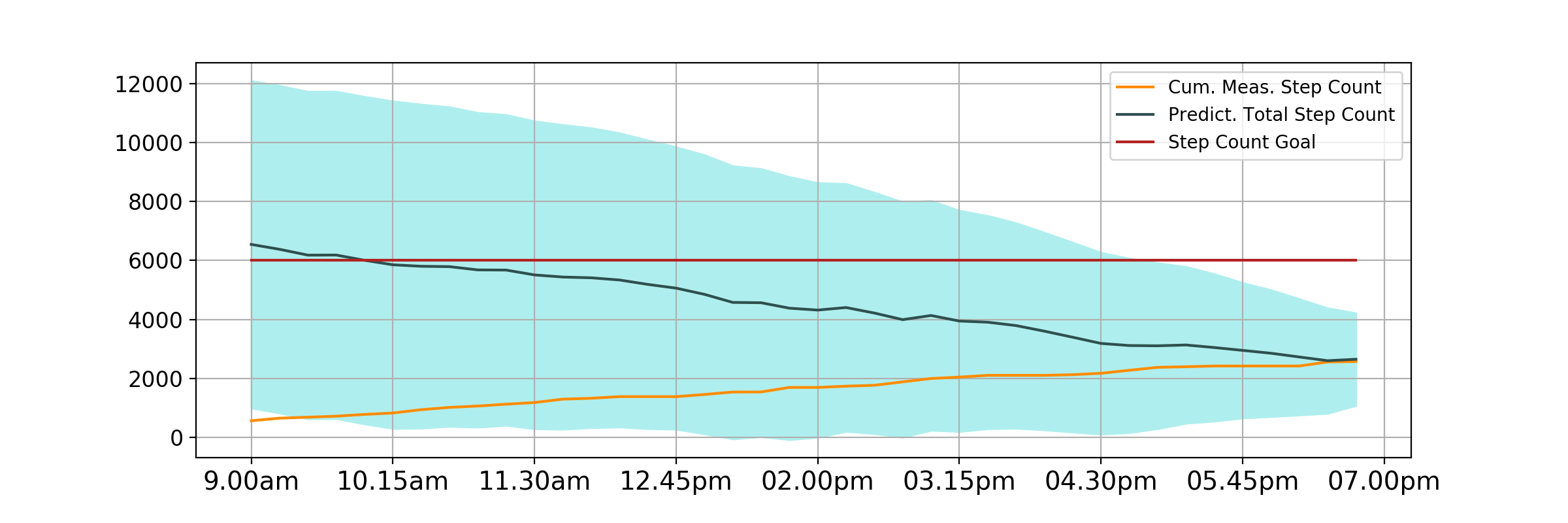}
		\caption{}
		\label{fig:detailed_weekday_low}
	\end{subfigure}
	\centering
	\caption{Text message interventions and the probability of achieving the desired goal for low physical activity. }
	\label{fig:low}
\end{figure}
\begin{figure}
	\centering
	\begin{subfigure}{0.9\linewidth}
		\centering
		\includegraphics[width=\linewidth]{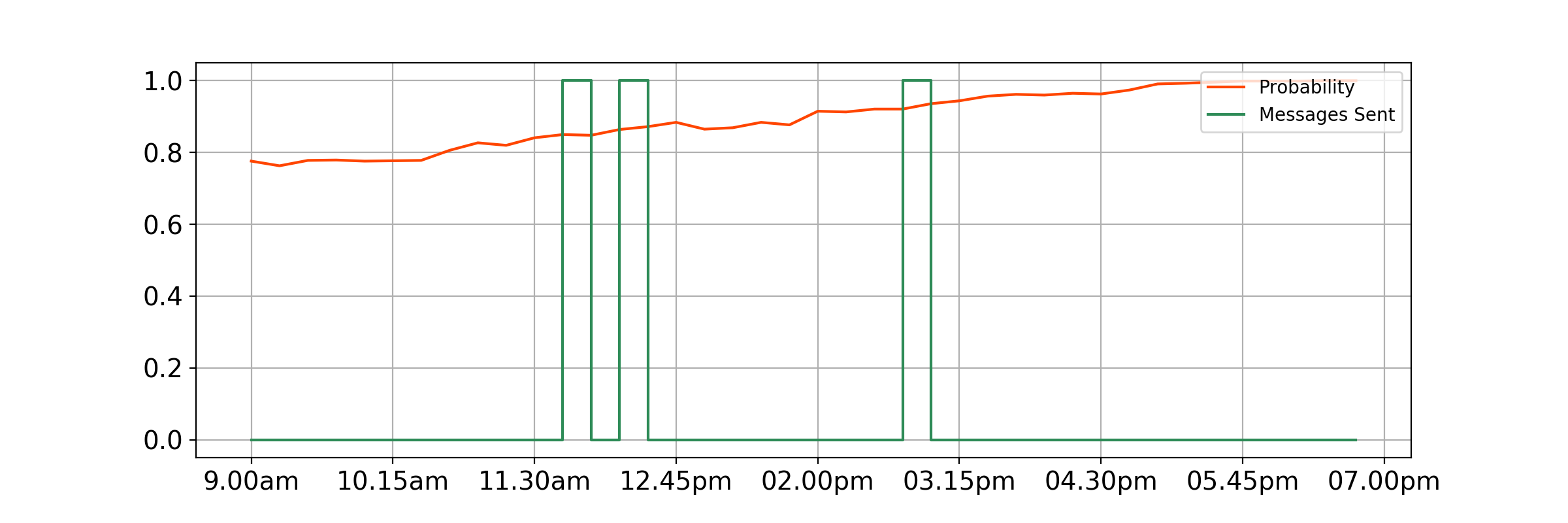}
		\caption{}
		\label{fig:input_weekday_high} 
	\end{subfigure}
	\hfill
	\begin{subfigure}{0.9\linewidth}
		\centering
		\includegraphics[width=\linewidth]{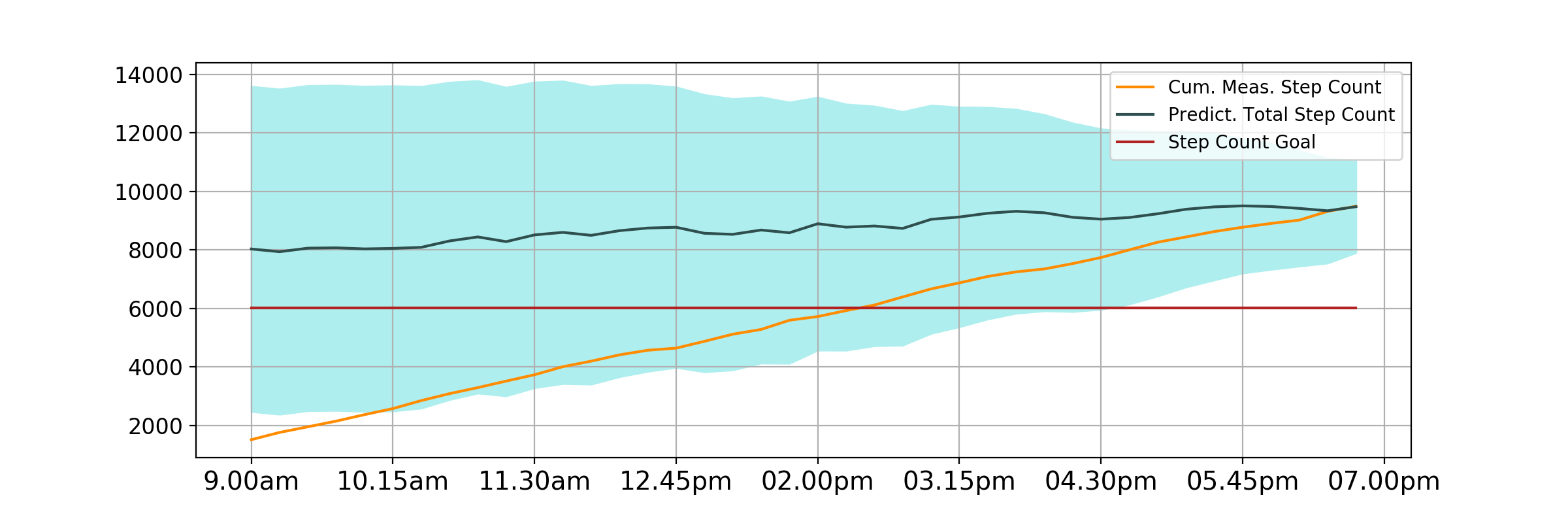}
		\caption{}
		\label{fig:detailed_weekday_high}
	\end{subfigure}
	\centering
	\caption{Text message interventions and the probability of achieving the desired goal for high physical activity.}
	\label{fig:high}
\end{figure}

In this application, the step count goal to be achieved is obtained by analyzing the physical activity of last 30 days. More specifically, the goal is set to average number of step counts taken in intervention window plus 500 steps for this period. For example, the intervention window for this subject is between 9:00am and 07:00pm. The average number of step count for this window is obtained as $ 5516 $ steps from the available data. Hence, the goal is set to $ 6016 $ steps. Furthermore, the measured step counts are generated using normal distribution with the mean $ \mu_{window} $ and standard deviation $ \sigma_{window} $ of the step counts in intervention window. These parameters are also estimated using the last 30 days of activity. In these simulations, the controller performance is analyzed for three different scenarios as regular, low and high volume physical activity, shown in figures \ref{fig:regular}, \ref{fig:low}, \ref{fig:high}, respectively.

\noindent \textbf{Scenario 1: Regular physical activity}\\
In this scenario, the measurements of step counts are generated using normal distribution with the mean $ \mu_{window} = 137 $ and standard deviation $ \sigma_{window}=51 $. 

\noindent \textbf{Scenario 2 \& 3: Low and high physical activity}\\
The measurements for low and high activity are generated using normal distribution with the mean $ 0.3 \mu_{window} $ and $ 1.5 \mu_{window} $ respectively, and standard deviation $ \sigma_{window} $ of the step counts in intervention window.

Figure \ref{fig:low}, that is related to low physical activity, shows that although the maximum number of messages allowed is set to 6, controller sends 4 messages. This can be interpreted as controller decides to stop sending messages since the probability of achieving the desired step count goal is so low later in the day.
Similarly, figure \ref{fig:high}, in the case of high physical activity, controller only sends 3 messages out of allowing 6. Again, this can be interpreted as controller decides to stop sending messages since the probability of achieving the desired step count goal is approaching to 1. Thus, this strategy optimizes treatment to achieve the minimum discrepancy between future activity and a desired step count (maximizing likelihood of goal attainment) with the fewest messages possible (minimizing burden).

\section{Conclusion}
This study contributes to design a model-based intervention MPC controller and an adaptation strategy that leads to implementing an efficient personalized adaptive messaging intervention for increasing physical activity, as well as reducing the burden of treatment. If successful, this strategy should help to sustain engagement and improve health and well-being.

\end{document}